\documentclass[aip,jcp,amssymb,amsmath,preprint]{revtex4-2}
\usepackage{amsmath}
\usepackage{cancel}
\usepackage{amsthm}
\usepackage{graphicx}
\usepackage{comment}

\def\be{\begin{equation}}
\def\ee{\end{equation}}
\def\ber{\begin{eqnarray}}
\def\eer{\end{eqnarray}}
\def\bern{\begin{eqnarray*}}
\def\eern{\end{eqnarray*}}

\def\rv{\mathbf{r}} %\def\rv{\ensuremath{\boldsymbol{r}}}

\def\Ev{\mathbf{E}}
\def\ev{\mathbf{e}}

\def\Jv{\mathbf{J}}

\def\Rv{\mathbf{R}} %\def\Rv{\ensuremath{\boldsymbol{R}}}

\def\vv{\mathbf{v}}

\def\0v{\mathbf{0}}
\def\1v{\mathbf{1}}
\def\2v{\mathbf{2}}
\def\3v{\mathbf{3}}

\def\av{\mathbf{a}}

\def\Acal{\mathbf{\mathcal{A}}}
\def\pa{\partial}

\DeclareMathAlphabet\mathbfcal{OMS}{cmsy}{b}{n}

\def\Re{ {\rm Re} \, }
\def\Im{ {\rm Im} \, }

\def\rvm{{\underset{\text{\raisebox{3 pt}{=}}}{r}}}

\begin{document}

\title{Stopping power of  electron liquid for slow quantum projectiles}
%\title{Stopping power of matter for quantum projectiles: Combined exact factorization and time-dependent density functional theory approach}
\author{Vladimir~U.~Nazarov}
\affiliation{Fritz Haber Research Center of Molecular Dynamics, the Hebrew University of Jerusalem, Institute of Chemistry, 9190401 Israel}
\email{vladimir.nazarov@mail.huji.ac.il}
\author{E.~K.~U.~Gross}
\affiliation{Fritz Haber Research Center of Molecular Dynamics, the Hebrew University of Jerusalem, Institute of Chemistry, 9190401 Israel}
%\email{eberhard.gross@mail.huji.ac.il}

\begin{abstract}
We revisit the problem of deceleration of a charge moving in a medium. 
Going beyond the traditional approach, which relies on Ehrenfest dynamics, we
treat the projectile fully quantum mechanically, on the same footing as the electrons of the target. 
In order to separate the dynamics of the projectile from that of the electrons, we employ the Exact Factorization method.
We illustrate the resulting  theory by applying it to the problem of the stopping power (SP) of a jellium-model metal for slowly moving charges. 
The quantum mechanical nature of particles  manifests itself remarkably in the differences in the SP for  projectiles of the same charge moving with the same velocity, but having different masses.
\end{abstract}

\maketitle

\section{Introduction}

Time-dependent density functional theory (TDDFT)\cite{Zangwill-80,Runge-84} in combination with Ehrenfest dynamics  is
a powerful method of the
theoretical study  of the stopping of ions in matter. 
Methodologically, it is useful to categorize several major regimes and first-principles approaches to the slowing of a charge in a medium:

(I) {\em High-velocity projectiles or weak projectile-target interaction}. 
In this case, the problem of the stopping power (SP) (the kinetic energy loss by a projectile per its unit path length) can be solved quite generally to the first Born approximation and be put in terms of the dielectric response of the target.  For a homogeneous medium, the solution is (see, e.g., Ref.~\onlinecite{Echenique-81})
\begin{equation}
S(v)\equiv -\frac{d E_{kin}}{d s}= \frac{2 Z^2}{\pi v^2} \int\limits_0^\infty \omega d \omega \int\limits_{\omega/v}^\infty \frac{d k}{k} \Im \left(-\frac{1}{\epsilon(k,\omega)} \right),
\label{S1B}
\end{equation}
where $E_{kin}$, $Z$, and $v$ are the kinetic energy, the charge, and the velocity of the projectile, respectively, 
and $\epsilon(k,\omega)$ is the wave-vector and frequency-dependent dielectric function  of the medium.
The problem remaining to be solved within this approach is the determination of 
$\epsilon(k,\omega)$, which is done with the use of the linear-response TDDFT, employing some of the available approximations to the exchange-correlation (xc) kernel $f_{xc}(k,\omega)$.\cite{Gross-85}  
While the first Born approximation severely limits the range of applicability of this approach, its advantage is that Eq.~(\ref{S1B}) can be readily generalized for arbitrary non-uniform target systems.

(II) {\em Low-velocity regime}. 
For the target of the homogeneous electron gas (HEG) (but only for this) the fully nonlinear solution of the SP problem for a slow projectile reads
\begin{equation}
Q_1=\lim_{v\to 0} \frac{S(v)}{v} =\bar n  \, k_F \sigma_{tr}(k_F),
\label{Q11}
\end{equation}
where $Q_1$ is known as the friction coefficient,  $\bar n$ is the density of the HEG, and $\sigma_{tr}(k_F)$ is the transport scattering cross-section of electron at the Fermi momentum level $k_F$. \cite{Finneman-68,Ferrell-77} While the potential at which the scattering in Eq.~(\ref{Q11}) occurs was 
originally approximated by some model,
it was further suggested to use the first-principles static Kohn-Sham\cite{Kohn-65} (KS) DFT potential  of a point charge immersed in the electron gas.\cite{Echenique-86}

(III) It was later  realized that Eq.~(\ref{Q11}) is incomplete, 
describing the binary-collisions (single-particle) contribution to the friction coefficient only (as indicated here by the subscript `1' at $Q$ in Eq.~(\ref{Q11})). Another contribution, which Eq.~(\ref{Q11}) misses, is the dynamic xc one \cite{Nazarov-05,Nazarov-07,Nazarov-08}
\begin{equation}
Q_2 = - \int [\hat{\vv}\cdot \nabla n_0(\rv)]
 \left. \frac{\pa \Im f_{xc}(\rv,\rv',\omega)}{\pa \omega}
\right|_{\omega=0} [\hat{\vv}\cdot \nabla' n_0(\rv')] d \rv \, d\rv',
\label{Q22}
\end{equation}
where $\hat{\vv}$ is the unit vector in the direction of the velocity, and $n_0(\rv)$ and $f_{xc}(\rv,\rv',\omega)$ are the  ground-state density and xc kernel, respectively, of the {\em inhomogeneous system of EG with the statically inserted projectile charge in it}.
The total friction coefficient is then given by 
\begin{equation}
Q=Q_1+Q_2.
\label{Q1122}
\end{equation}
We note that Eqs.~(\ref{Q11})-(\ref{Q1122}) are formally exact within the low-velocity regime, while approximations are invoked 
in calculations, when choosing specific functionals for the static KS potential $v_{KS}(\rv)$ and the dynamic xc kernel $f_{xc}(\rv,\rv',\omega)$.

(IV) The problem of the SP can, quite generally, be treated by means of real-time TDDFT. 
This approach  neither imposes limitations  on the type of the target system nor on the velocity of the ion, and it has now become a standard first-principles technique for large-scale SP calculations in crystals.\cite{Baer-04,Pruneda-07}

All the approaches briefly reviewed above consider a projectile moving as a  point charge according to the laws of classical mechanics. 
Since there exist no classical particles in nature, the purpose of this work is to develop a theory of SP taking into account quantum mechanical effects on the part of projectiles, and to analyse their comparative importance. 
Specifically, within the traditional classical approach, for given charge and velocity of a projectile, the SP, obviously, does not depend on the mass of the latter. We will see that this is not the case within the quantum treatment.

The organization of this paper is the following. 
In Sec.~\ref{formalism}, based on the Exact Factorization (EF) method, we develop a theory of SP for a quantum projectile.
To illustrate the general theory by a simple calculation,
in Sec.~\ref{QQEE} we invoke the approximation of the uncorrelated projectile-electrons motion (mean field approximation).
The latter approximation allows for the reformulation of the quantum SP problem in terms of TDDFT, which is done in Sec.~\ref{TDDFT}.
In Sec.~\ref{low} we focus on the low-velocity SP, developing the linear-response TDDFT approach to this problem
(our small parameter is the velocity, not the projectile-target interaction!).
In Sec.~\ref{ress} we present and discuss results of calculations. Section \ref{concl} contains conclusions.
Lengthy derivations are moved to Appendices. 

\section{Stopping power problem with a quantum projectile: Exact Factorization approach}
\label{formalism}
For the sake of maximal clarity, we consider a jellium model of a metal with a fixed positive charge background density $|e| \bar{n}$. 
\footnote{The generalization for crystals with frozen lattices is straightforward. The generalization for solids with account of the quantum nature of nuclei comprising the lattices would be challenging, and it will be a subject of separate studies.}
A projectile particle with charge $|e| Z$, which position vector we denote by $\Rv$ and which is distinguishable from electrons, is traversing this system.
Our unperturbed Hamiltonian is
\begin{equation}
\hat{H}_0 =-\frac{\hbar^2}{2 M} \nabla_\Rv^2+\hat{H}_{BO},
\label{H000}
\end{equation}
where $M$ is the mass of the projectile and $\hat{H}_{BO}$ is the Born–Oppenheimer Hamiltonian 
\begin{equation}
\begin{split}
\hat{H}_{BO} &= \sum\limits_{i=1}^{N} \left[ -\frac{\hbar^2}{2 m} \nabla_{\rv_i}^2 - \frac{Z e^2}{|\Rv-\rv_i|} 
-  \int \frac{e^2 \bar{n}}{|\rv_i-\rv|} d\rv\right] \\
&+\frac{1}{2} \sum_{i\ne j}^{N} \frac{e^2}{|\rv_i-\rv_j|}
+  \int \frac{Z e^2 \bar{n}}{|\Rv-\rv|} d\rv  + \frac{1}{2} \int \frac{e^2 \bar{n}^2}{|\rv-\rv'|} d\rv d\rv',
\end{split}
\label{BO}
\end{equation}
where $m$ is the electron mass  and $\rv_i$ is the position of the $i$-th electron.
Following the Exact Factorization (EF) formalism,\cite{Abedi-10}
we represent the many-body wave-function $\Psi$ of the system  as  
\begin{equation}
\Psi(\Rv,\rvm,t)=\chi(\Rv,t) \Phi_\Rv(\rvm,t),
\label{EFexact}
\end{equation}
where by $\rvm$ we denote the set of the coordinates of all electrons.
The normalization conditions are imposed
\begin{align*}
&\langle \chi(\Rv,t)|\chi(\Rv,t)\rangle_\Rv=1, \ \forall \, t\\
&\langle \Phi_\Rv(\rvm,t)|\Phi_\Rv(\rvm,t)\rangle_\rvm=1, \ \forall \, \Rv, t,
\end{align*}
where  $\langle \dots \rangle_\Rv$ and $\langle \dots \rangle_\rvm$ denote the integration over the corresponding coordinates.
The `wave-function'
$\chi(\Rv,t)$ obeys the equation of motion\cite{Abedi-10}
\begin{align}
&i\hbar \pa_t\chi(\Rv,t)= \hat{H}_\chi(t) \, \chi(\Rv,t),\\
&\hat{H}_\chi(t)= \frac{1}{2 M} \left[ -i \hbar \nabla_\Rv+ \Acal(\Rv,t)\right]^2 +\epsilon(\Rv,t),
\end{align}
where
\begin{align}
\Acal(\Rv,t) &=-i\hbar \langle \Phi_\Rv(\rvm,t)|\nabla_\Rv|\Phi_\Rv(\rvm,t)\rangle_\rvm \label{Aequ},\\
\begin{split}
\epsilon(\Rv,t) &=\langle \Phi_\Rv(\rvm,t) |\hat{H}_{BO}| \Phi_\Rv(\rvm,t) \rangle_\rvm +G(\Rv,t) 
-i \hbar  \langle \Phi_\Rv(\rvm,t)|\pa_t \Phi_\Rv(\rvm,t)\rangle_\rvm \\
&+V^{(n)}_{ext}(\Rv,t),
\end{split}\\
G(\Rv,t) &=\frac{\hbar^2}{2 M}    \langle \nabla_\Rv\Phi_\Rv(\rvm,t)|\nabla_\Rv\Phi_\Rv(\rvm,t)\rangle_\rvm-
\frac{1}{2 M} \Acal^2(\Rv,t) \label{Gequ},
\end{align}
and where  $V^{(n)}_{ext}(\Rv,t)$ is a potential, possibly applied to the projectile externally.

We use the fact that the exact rate of change of the kinetic energy of the projectile is given by\cite{Li-22}
\begin{equation}
\frac{d E_{kin}(t)}{d t} = 
\int \Jv^{(n)}(\Rv,t)\cdot \left[ \pa_t \Acal(\Rv,t) -\nabla_\Rv \epsilon(\Rv,t)\right] d\Rv +
\frac{d}{dt} \int \Gamma^{(n)}(\Rv,t) G(\Rv,t) d\Rv,
\label{dEkindt}
\end{equation}
where $\Gamma^{(n)}(\Rv,t)$ and $\Jv^{(n)}(\Rv,t)$ are the projectile's particle density and current-density, respectively,
\begin{align}
&\Gamma^{(n)}(\Rv,t)= |\chi(\Rv,t)|^2,\\
&\Jv^{(n)}(\Rv,t)=\frac{\hbar}{M} {\rm Im}\, [\chi^*(\Rv,t) \nabla_\Rv \chi(\Rv,t)]+\frac{1}{M} \Acal(\Rv,t) \Gamma^{(n)}(\Rv,t).
\label{jnot}
\end{align}

Equation (\ref{dEkindt}) is a general and exact result for the change of the kinetic energy of a particle interacting with other particles of a system but singled out within the paradigm of EF. 
The second term in Eq.~(\ref{dEkindt}) was found to be associated with a geometric energy transfer between electrons and nuclei.\cite{Requist-22}
To find the quantities $\Acal(\Rv,t)$, $G(\Rv,t)$, $\epsilon(\Rv,t)$ entering this equation requires the knowledge of the conditional electronic wave-function $\Phi_\Rv(\rvm,t)$. 
Since the exact determination of  $\Phi_\Rv(\rvm,t)$
is an insurmountable task equivalent to the solution of the original many body problem, in practice the use of approximations is necessary.
In order to illustrate our theory with the simplest possible non-trivial example, in the next section we resort to a mean field approximation. 

\section{Mean field approximation}
\label{QQEE}

We use the approximation to Eq.~(\ref{EFexact})
\begin{equation}
\Psi(\Rv,\rvm,t)=\chi(\Rv,t) \Phi(\rvm,t),
\label{EF0}
\end{equation}
where we neglect the $\Rv$-dependence of the conditional electronic wave-function $\Phi_\Rv(\rvm,t)$.
With the use of the time-dependent (TD) variational principle of Ref.~\onlinecite{McLachlan-64}, in Appendix \ref{TDvarapp} we show
that the optimal $\chi(\Rv,t)$ and $\Phi(\rvm,t)$, 
i.e., such that the many-body 
TD Schr\"{o}dinger equation is satisfied to the best accuracy compatible with Eq.~(\ref{EF0}), obey equations of motion:

\noindent{\em Projectile motion}:
\begin{align}
&i \hbar \pa_t \chi(\Rv,t) = [\hat{H}^{(n)}+\bar{V}^{(n)}(\Rv,t)] \chi(\Rv,t), \label{N1} \\
&\hat{H}^{(n)}=-\frac{\hbar^2}{2 M} \nabla_\Rv^2+ Z \int \frac{e^2 \bar{n}}{|\Rv-\rv|} d\rv +V^{(n)}_{ext}(\Rv,t) ,\\
&\bar{V}^{(n)}(\Rv,t)=\langle \Phi(\rvm,t)|\hat{H}_{en} |\Phi(\rvm,t) \rangle_\rvm=-Z \int \frac{e^2 n^{(e)}(\rv,t)}{|\Rv-\rv|} d\rv \label{N3},\\
&\hat{H}_{en}= -Z \sum\limits_{i=1}^{N_e} \frac{e^2 }{|\Rv-\rv_i|}.
\end{align}

\noindent{\em Electron motion}:
\begin{align}
&i \hbar \pa_t \Phi(\rvm,t)   = [\hat{H}^{(e)}+\bar{V}^{(e)}(\rvm,t)] \Phi(\rvm,t), \label{E1}\\
&\hat{H}^{(e)}=  \! \sum\limits_{i=1}^{N_e} \left[ -\frac{\hbar^2}{2 m} \nabla_{\rv_i}^2 
\! - \! \int \frac{e^2 \bar{n}}{|\rv_i-\rv|} d\rv \right]+\frac{1}{2} \sum_{i\ne j}^{N_e} \frac{e^2}{|\rv_i-\rv_j|} + \frac{1}{2} \int \frac{e^2 \bar{n}^2}{|\rv-\rv'|} d\rv d\rv',\\
& \bar{V}^{(e)}(\rvm,t)=\langle \chi(\Rv,t) |\hat{H}_{en} |\chi(\Rv,t) \rangle_\Rv=-Z \sum\limits_{i=1}^{N_e} \int \frac{e^2 n^{(n)}(\Rv,t)}{|\Rv-\rv_i|} d\Rv
\label{E3}.
\end{align}
Theory based on Eqs.~(\ref{N1})-(\ref{E3})
has been known before as the mean-field time-dependent self-consistent field (TDSCF) method.
\footnote{ Equations~(\ref{N1})-(\ref{E3}) were obtained in Refs.~\onlinecite{Gerber-82,Tully-98} in a different way.
We, however, give the preference to their derivation from the TD variational principle,
since it reveals the place of TDSCF method within the framework of the Exact Factorization.}

We see that, within the approximation (\ref{EF0}), the electrons move in the potential of the distributed charge density of the projectile and vice versa. It is also immediately seen that in the limit of the infinitesimally narrow projectile wave-packet $|\chi(\Rv,t)|^2\to \delta(\Rv-\Rv(t))$
the classical Ehrenfest dynamics is reproduced. 
Furthermore, by Eqs.~(\ref{Aequ}) and (\ref{Gequ}),
\begin{equation}
\Acal(\Rv,t)=\0v, \ \ G(\Rv,t)=0,
\label{AQ0}
\end{equation}
and, by Eqs.~(\ref{BO}), (\ref{E1})-(\ref{E3}),
\begin{equation}
\epsilon(\Rv,t)=\int \frac{Z  e^2 [\bar{n}-n^{(e)}(\rv,t)]}{|\Rv-\rv|} d\rv 
+V^{(n)}_{ext}(\Rv,t).
\label{eps0}
\end{equation}

With account of Eqs.~(\ref{AQ0}) and (\ref{eps0}), and with the use of the continuity equation for the projectile,  Eq.~(\ref{dEkindt})
reduces to
\begin{equation}
\frac{d E_{kin}(t)}{d t} = 
-\int \epsilon(\Rv,t)  \pa_t n^{(n)}(\Rv,t) d\Rv .
\label{dEkindt0}
\end{equation}

In conclusion of this section we note that the approximation~(\ref{EF0}) can be considered as the zeroth-order term of a perturbation  series in powers of $\hat{H}_{en}(\Rv,\rvm)-\bar{V}^{(n)}(\Rv,t)-\bar{V}^{(e)}(\rvm,t)$. In this expansion, the 1st-order contribution to the kinetic energy loss in Eq.~(\ref{dEkindt}) can be shown to vanish identically.

\section{TDDFT reformulation}
\label{TDDFT}

The electronic motion problem (\ref{E1})-(\ref{E3}) can be reduced to that of TDDFT. 
Indeed, the time-dependent KS equations for electronic orbitals can be written as
\begin{equation}
\begin{split}
i \hbar \pa_t \phi_i(\rv,t)   & = 
\left[-\frac{\hbar^2}{2 m} \nabla_{\rv}^2
 +   \int  \frac{e^2 [n^{(e)}(\rv',t)  -  \bar{n}]}{|\rv-\rv'|} d\rv'  +  V_{xc}^{(e)}(\rv,t)  \right. \\
 & \left. -  Z   \int \frac{e^2 n^{(n)}(\Rv,t)}{|\rv-\Rv|} d\Rv  +  V^{(e)}_{ext}(\rv,t)\right]  \phi_i(\rv,t),
\end{split} 
\end{equation}
\begin{equation}
n^{(e)}(\rv,t)= \sum\limits_{i=1}^{N_e} |\phi_i(\rv,t)|^2,
\end{equation}
and solved mutually-consistently  with the projectile's motion problem (\ref{N1})-(\ref{N3}).

\subsection{Ground state}
\label{GS}
As the initial condition, in the following we will need the ground-state solution for the system of the projectile at rest in the electron gas.
We solve the electronic ground-state DFT problem
\begin{align}
&\left[-\frac{\hbar^2}{2 m} \nabla_{\rv}^2
\! + \! \! \int \! \frac{e^2 [n^{(e)}(\rv') \! - \! \bar{n}]}{|\rv-\rv'|} d\rv' \! + \! V_{xc}^{(e)}(\rv) \! - \! Z \! \! \int \frac{e^2 n^{(n)}(\Rv)}{|\rv-\Rv|} d\Rv \right]  \phi_i(\rv)=\epsilon_i \phi_i(\rv),\\
&n^{(e)}_0(\rv)= \sum\limits_{i=1}^{N_e} |\phi_i(\rv)|^2,
\end{align}
together with the corresponding nuclear problem
\begin{align}
&\left[-\frac{\hbar^2}{2 M} \nabla_\Rv^2+ V^{(n)}_0(\Rv) \right] \chi_0(\Rv)=E_0 \chi_0(\Rv),\\
&V^{(n)}_0(\Rv)= -Z \int \frac{e^2 [n^{(e)}_0(\rv)-\bar{n}]}{|\Rv-\rv|} d\rv.
\end{align}

\begin{figure*}[h!]
\includegraphics[clip=true, trim= 30 0 0 0, width=1 \textwidth]{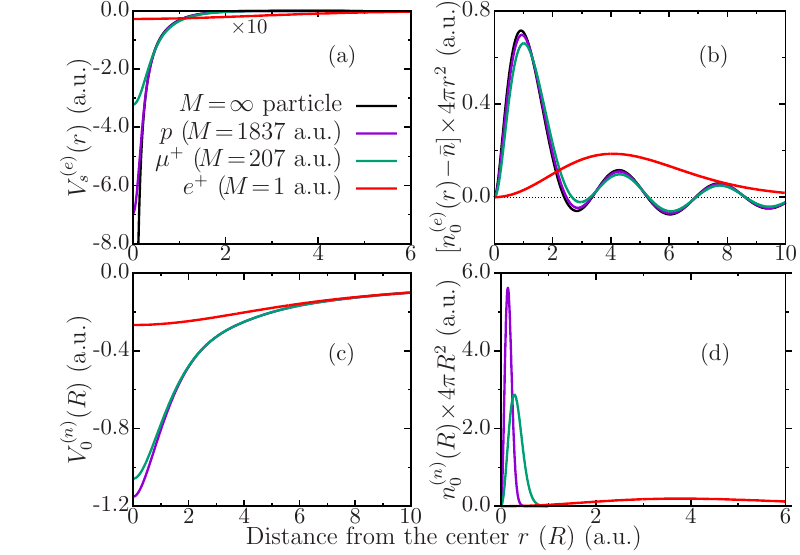}
\caption{\label{static}
Self- and mutually consistent ground-state properties of the system of the impurity particle of the charge $Z=+1$ a.u. immersed in the jellium-model electron gas of the density parameter $r_s=2.07$. The cases of the infinite mass particle, proton ($M=1837$ a.u.), antimuon ($M=207$ a.u.), and positron ($M=1$ a.u.) are compared.
(a) KS potential of the electronic system; (b) The electron density relative to that of the unperturbed HEG; (c) The potential experienced by  the impurity particle and (d) the corresponding impurity particle density distribution. }
\end{figure*}

\begin{figure*}[h!]
\includegraphics[clip=true, trim= 30 0 0 0, width=1 \textwidth]{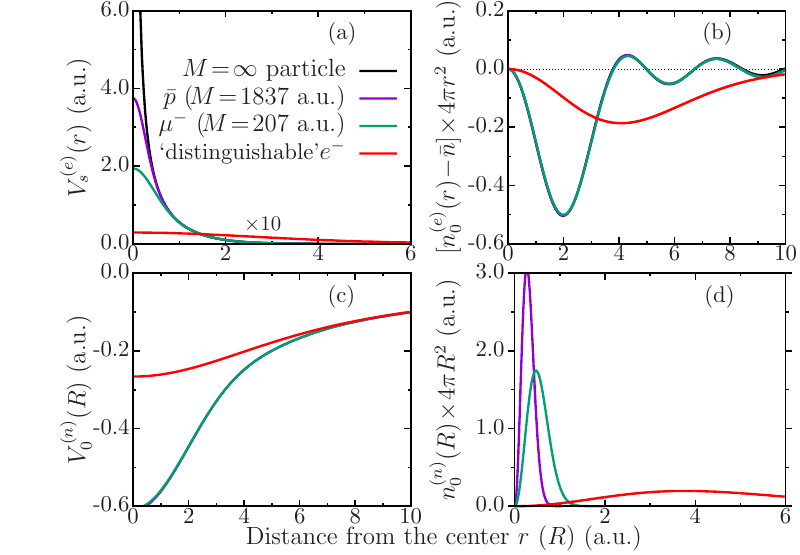}
\caption{\label{staticn} The same as Fig.~\ref{static} but for  negatively charged impurities of $Z=-1$ a.u.
The cases of the infinite mass particle, antiproton, muon, and a fictitious distinguishable electron are compared.
Some lines are too close to each other to be discerned. 
}
\end{figure*}

Results of example calculations of the ground-state potentials and the electronic and nuclear densities distributions are presented in Figs.~\ref{static} and \ref{staticn}. We point out that a bound state is formed in both cases of the positively and negatively charged nuclei. 
This is due to the presence of the uniform positive background charge density, 
ensuring the depletion (accumulation) of the net positive charge in the vicinity of the positive (negative) nucleus, leading in both cases to the confinement of the latter.


\section{Low-velocity stopping power}
\label{low}

Below we focus on slowly moving  projectiles, in which regime more significant quantum effects can be expected.

\subsection{Method}

To make the charge move,  we apply to it a weak  external potential
\begin{equation}
V^{(n)}_{ext}(\Rv,t)= -(\Ev\cdot\Rv) \cos(\omega t).
\label{Vnext}
\end{equation}
To avoid the difficulties of the steady-state approach, this potential is taken time-dependent monochromatic.\cite{Nazarov-14-2} 
We solve the problem of SP perturbatively in powers of $\Ev$ and, consequently, in powers of the velocity. 
In the end, to relate our results to the constant velocity regime, 
we take the $\omega\to 0$ limit.
In Eq.~(\ref{dEkindt0}),  we find that the 1st order contribution in $\Ev$ is zero identically
\begin{equation}
\left(\frac{d E_{kin}(t)}{d t}\right)_1 =
- \int V^{(n)}_0(\Rv) \pa_t n^{(n)}_1(\Rv,t) d\Rv =0,
\end{equation}
(subscripts indicate orders in the expansion of the corresponding quantities in powers of $\Ev$).
Therefore, we are concerned with  the 2nd order contribution
\begin{equation}
\left(\frac{d E_{kin}(t)}{d t}\right)_2 =
- \int V^{(n)}_0(\Rv) \pa_t n^{(n)}_2(\Rv,t) d\Rv -\int  V^{(n)}_1(\Rv,t) \pa_t n^{(n)}_1(\Rv,t) d\Rv,
\label{dEkindt2}
\end{equation}
\begin{equation}
V^{(n)}_1(\Rv,t)=-(\Ev\cdot\Rv) \cos(\omega t)-Z \int \frac{ e^2 n^{(e)}_1(\rv,t)}{|\Rv-\rv|} d\rv,
\label{ggg}
\end{equation}
where the 1st term on the RHS of Eq.~(\ref{ggg}) is due to the externally applied potential (\ref{Vnext}) and the 2nd one is 
the potential of the dynamically perturbed electronic density.
In Appendix \ref{1st} it is shown that the expectation value of the instantaneous velocity of the projectile is
\begin{equation}
\vv(t)= \vv \cos(\omega  t),
\label{veq}
\end{equation}
where $\vv$ is related to $\Ev$ through the system of three equations
\begin{equation}
\sum_{j=1}^3 v_j \int \nabla_i n^{(n)}_0(\Rv) \left .\pa_\omega \Im \Pi(\Rv,\Rv',\omega)\right|_{\omega=0} \nabla'_j n^{(n)}_0(\Rv') d\Rv d\Rv' = - E_i,
\label{veqeq}
\end{equation}
where
\begin{equation}
\Pi(\Rv,\Rv',\omega)= Z^2 \int  \frac{ \chi^{(e)}_1(\rv,\rv',\omega)}{|\Rv-\rv| |\Rv'-\rv'|} d\rv d \rv',
\label{Pieq}
\end{equation}
$n^{(n)}_0(\Rv)$ is the ground-state density distribution of the projectile, and $\chi^{(e)}_1(\rv,\rv',\omega)$ is the linear density response function of the electronic system w.r.t. its ground state.
The friction coefficient can be written as
\begin{equation}
Q=- \frac{1}{v(t)} \left(\frac{d E_{kin}(t)}{d s}\right)_2=- \frac{1}{v^2(t)} \left(\frac{d E_{kin}(t)}{d t}\right)_2.
\label{Qprel}
\end{equation}
Our evaluation of the last expression on the RHS of Eq.~(\ref{Qprel}) is based on Eqs.~(\ref{dEkindt2}) and (\ref{veq})-(\ref{Pieq}). 
The crucial points leading to the success of this method are: 
\begin{enumerate}
\item
Although the 1st term on the RHS of Eq.~(\ref{dEkindt2})
includes $n^{(n)}_2(\Rv,t)$ and, apparently, involves the 2nd order response functions, the latter disappear from the final result, with only the 1st order response functions left. It is shown in Appendix \ref{a2dn} that this becomes possible due to the sum rules (\ref{sr1}), (\ref{sr2})-(\ref{sr3}), which the 2nd order density response function obeys.
\item
In the $\omega\to 0$ limit (taken only outside the trigonometric functions of the argument $\omega t$, but not inside them, because $t$ can be large), 
the latter functions cancel out in Eq.~(\ref{Qprel}), leading to the constant friction coefficient, not depending  on the instantaneous velocity.
\end{enumerate} 
The above properties are proved in Appendix \ref{mainder}, where the following expression 
for the friction coefficient is derived
\begin{equation}
Q  =
  \int (\hat{\ev}\cdot \nabla) V^{(e)}_0(\rv) \left .\pa_\omega \Im \chi^{(e)}_1(\rv,\rv',\omega)\right|_{\omega=0} (\hat{\ev}\cdot \nabla') V^{(e)}_0(\rv') d\rv d\rv',
\label{res}
\end{equation}
where $\hat{\ev}$ is the unit vector parallel to the driving electric field $\Ev$ (or to the velocity). 

Equation (\ref{res}) differs from the corresponding result of Ref.~\onlinecite{Nazarov-05} for a classical projectile  by $V^{(e)}_0(\rv)$ in place of the point charge Coulomb potential [both explicitly and affecting $\chi^{(e)}_1(\rv,\rv',\omega)$].
Accordingly, \cite{Nazarov-05}  we can rewrite Eq.~(\ref{res}) in terms of the linear response TDDFT quantities as 
\begin{align}
\begin{split}
Q_1 &= - \int [\hat {\ev} \cdot \nabla V^{(e)}_s(\rv) ]
\left. \pa_\omega \Im \chi^{(e)}_s(\rv,\rv',\omega) \right|_{\omega=0} [\hat{\ev} \cdot \nabla' V^{(e)}_s(\rv') ] d\rv  d\rv' ,
\label{Q1} 
\end{split} \\
Q_2 &= - \int [\hat{\ev}\cdot \nabla n^{(e)}_0(\rv)]
 \left. \pa_\omega \Im f^{(e)}_{xc}(\rv,\rv',\omega)
\right|_{\omega=0} [\hat{\ev}\cdot \nabla' n^{(e)}_0(\rv')] d \rv \, d\rv',
\label{Q2}
\end{align}
\begin{equation}
Q  =   Q_1  +  Q_2,
\label{Q1Q2}
\end{equation}
where $\chi^{(e)}_s$ and $V^{(e)}_s$ are the Kohn-Sham (KS) density response function and the static KS potential, respectively, of the electronic system, $f^{(e)}_{xc}$ is the corresponding xc kernel,  
and $n^{(e)}_0$ is the electronic ground-state density. As proved in Ref.~\onlinecite{Nazarov-05}, Eq.~(\ref{Q1}) is equivalent to the binary-collision part of the friction coefficient (\ref{Q11}).

\section{Results and discussion}
\label{ress}

\begin{figure*}[h]
\includegraphics[clip=true, trim= 85 0 0 0, width=1 \textwidth]{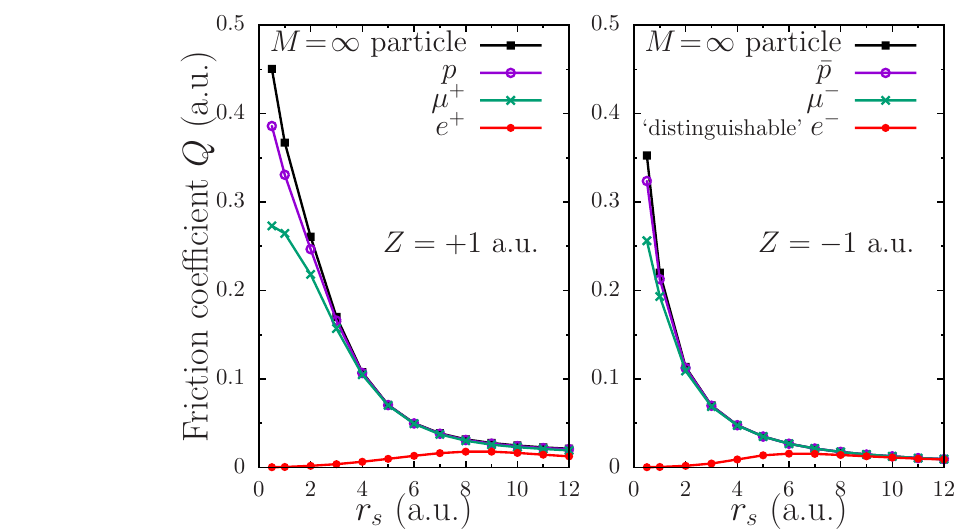}
\caption{\label{q1q2sp_and_n}
Friction coefficient versus the electron gas density parameter $r_s$ for positively (left) and negatively (right) charged projectiles. }
\end{figure*}

We have conducted calculations by formulas (\ref{Q11}), (\ref{Q2}), and (\ref{Q1Q2}), using the local density approximation (LDA) for the static xc potential $V^{(e)}_{xc}(\rv)$ and LDA frequency-dependent  xc kernel $f^{(e)}_{xc}(\rv,\rv',\omega)$.\cite{Gross-85}
In Fig.~\ref{q1q2sp_and_n}, results of calculations of the friction coefficient for projectiles of the charge of $Z=+1$ and $Z=-1$ and various masses are presented. The curve marked as $M=\infty$ shows the friction of the classical particle.
For massive projectiles, at higher EG densities (smaller $r_s$), we observe appreciable differences in the friction coefficient for particles of the same charge but different masses, the latter being a specifically quantum-mechanical effect on the part of the projectiles. 

For light projectiles (positron and fictitious distinguishable electron) quantum theory results differ drastically from the classical ones. At smaller $r_s$ (high EG densities) the friction is almost totally suppressed. It grows with the growing $r_s$, developing a maximum (around $r_s\approx 8$ and $7$, for positron and `electron', respectively). 
This behaviour of $Q$ has nothing in common with that of the classical particle, the latter shown by the curve with $M=\infty$.
It can be easily realized that the mass dependence of the friction coefficient is a consequence of the differences in the wave-packets of projectiles with different masses, and this effect does not have a classical analogue.

\section{Conclusions}
\label{concl}

Within the Exact Factorization approach, we have developed a theory of the stopping of charges moving in media, 
when not only electrons of a medium, but also the projectile itself is treated quantum mechanically.
Using a simple approximation  of
the mean-field time-dependent self-consistent field method (the uncorrelated projectile-electrons dynamics),
we have illustrated our theory by calculations of the friction coefficient for slowly moving charges in jellium-model metal. 

We have identified a principle effect of the quantum mechanical approach to the stopping power problem: particles bearing the same charge and moving with the same velocity experience different friction depending on their respective masses. 
The latter is a result of the differences in the wave packets' sizes of particles with different masses, and this effect, obviously, cannot arise
within the classical Ehrenfest dynamics.

\

\acknowledgements
This project has received funding from the European Research Council (ERC) under the European Union’s Horizon 2020 research and innovation programme (Grant Agreement No. ERC-2017-AdG-788890). E.K.U.G. acknowledges support as Mercator Fellow at the University Duisburg-Essen within SFB 1242 funded by the Deutsche Forschungsgemeinschaft (DFG, German Research Foundation) Project No. 278162697.

\

\noindent
The authors declare no conflicts of interest.

\

\noindent
The data that support the findings of this study are available from the authors upon reasonable request.

\appendix

\section{Mean field approximation: Derivation of Eqs.~(\ref{N1})-(\ref{E3})}
\label{TDvarapp}

Using the TD variational principle of McLachlan,\cite{McLachlan-64} we minimize the functional
\begin{equation}
F=\int \left|  i \hbar \pa_t \Psi(\Rv,\rvm,t) - \hat{H}(t) \Psi(\Rv,\rvm,t) \right|^2 d\Rv d\rvm,
\end{equation}
where the trial function $\Psi$ is restricted to the form (\ref{EF0}), 
\begin{equation}
\hat{H}(t)=\hat{H}_0+V^{(n)}_{ext}(\Rv,t),
\label{HHHH}
\end{equation}
and $\hat{H}_0$ is given by Eq.~(\ref{H000}). At each time moment $t$, we vary the time-derivatives of  $\chi(\Rv,t)$ and $\Phi(\rvm,t)$, while these functions themselves are considered fixed.\cite{McLachlan-64} Therefore,
\begin{equation}
\begin{split}
0=\delta F= &-2 \hbar\Re i  \int \left[  i \hbar \pa_t \Psi^*(\Rv,\rvm,t) + \hat{H}(t) \Psi^*(\Rv,\rvm,t) \right]   \Phi(\rvm,t) \delta \pa_t \chi(\Rv,t) d\Rv d\rvm \\
&-2 \hbar\Re i  \int \left[  i \hbar \pa_t \Psi^*(\Rv,\rvm,t) + \hat{H}(t) \Psi^*(\Rv,\rvm,t) \right]   \chi(\Rv,t)  \delta \pa_t \Phi(\rvm,t) d\Rv d\rvm.
\end{split}
\end{equation}
Then
\begin{equation}
\begin{split}
&\Im \int \left[  i \hbar \pa_t \Psi^*(\Rv,\rvm,t) + \hat{H}(t) \Psi^*(\Rv,\rvm,t) \right]   \Phi(\rvm,t) \delta \pa_t \chi(\Rv,t) d\Rv d\rvm \\
&+ \Im \int \left[  i \hbar \pa_t \Psi^*(\Rv,\rvm,t) + \hat{H}(t) \Psi^*(\Rv,\rvm,t) \right]   \chi(\Rv,t)  \delta \pa_t \Phi(\rvm,t) d\Rv d\rvm =0,
\end{split}
\end{equation}
or
\begin{equation}
\begin{split}
& \int \Im \left\{\left[  i \hbar \pa_t \Psi^*(\Rv,\rvm,t) + \hat{H}(t) \Psi^*(\Rv,\rvm,t) \right]   \Phi(\rvm,t)\right\} \Re \delta \pa_t  \chi(\Rv,t) d\Rv d\rvm \\
& +\int \Re \left\{\left[  i \hbar \pa_t \Psi^*(\Rv,\rvm,t) + \hat{H}(t) \Psi^*(\Rv,\rvm,t) \right]   \Phi(\rvm,t)\right\} \Im \delta \pa_t \chi(\Rv,t) d\Rv d\rvm \\
&+  \int \Im  \left\{ \left[  i \hbar \pa_t \Psi^*(\Rv,\rvm,t) + \hat{H}(t) \Psi^*(\Rv,\rvm,t) \right]   \chi(\Rv,t) \right\}  \Re \delta \pa_t  \Phi(\rvm,t) d\Rv d\rvm \\
&+  \int \Re  \left\{ \left[  i \hbar \pa_t \Psi^*(\Rv,\rvm,t) + \hat{H}(t) \Psi^*(\Rv,\rvm,t) \right]   \chi(\Rv,t) \right\}  \Im \delta \pa_t  \Phi(\rvm,t) d\Rv d\rvm =0.
\end{split}
\label{B5}
\end{equation}
Since both real and imaginary parts of both $\delta \pa_t \chi(\Rv,t)$ and $\delta \pa_t  \Phi(\rvm,t)$ are arbitrary functions of their respective arguments, 
from Eq.~(\ref{B5}) we conclude
\begin{align}
& \int \left[  i \hbar \pa_t \Psi^*(\Rv,\rvm,t) + \hat{H}(t) \Psi^*(\Rv,\rvm,t) \right]   \Phi(\rvm,t) d\rvm =0,  \\
& \int  \left[  i \hbar \pa_t \Psi^*(\Rv,\rvm,t) + \hat{H}(t) \Psi^*(\Rv,\rvm,t) \right]   \chi(\Rv,t)  d\Rv =0. 
\end{align}
Taking the complex conjugate of the above equations, we can write with the use of Eq.~(\ref{EF0})
\begin{align}
&  -i \hbar \pa_t \chi(\Rv,t) + \langle \Phi(\rvm,t)|\hat{H}(t) | \chi(\Rv,t) \Phi(\rvm,t)\rangle_\rvm   
-i \hbar \langle \Phi(\rvm,t)| \pa_t \Phi(\rvm,t)\rangle_\rvm \chi(\Rv,t)=0,  \\
& -i \hbar \pa_t \Phi(\rvm,t) +\langle   \chi(\Rv,t)| \hat{H}(t)| \chi(\Rv,t) \Phi(\rvm,t)\rangle_\Rv   -i \hbar \langle \chi(\Rv,t)| \pa_t \chi(\Rv,t)\rangle_\Rv \Phi(\rvm,t)   =0. 
\end{align}
With account of Eqs.~(\ref{HHHH}) and (\ref{H000}), and (\ref{BO}), the latter can be written as
\begin{align}
&  i \hbar \pa_t \chi(\Rv,t) = [\hat{H}^{(n)}(t)+\bar{V}^{(n)}(\Rv,t)+C^{(n)}(t)] \chi(\Rv,t),  \\
& i \hbar \pa_t \Phi(\rvm,t) =[\hat{H}^{(e)}+\bar{V}^{(e)}(\rvm,t)+C^{(e)}(t)] \Phi(\rvm,t), 
\end{align}
where  $\hat{H}^{(n)}(t)$, $\hat{H}^{(e)}$, $\bar{V}^{(n)}(\Rv,t)$ and $\bar{V}^{(e)}(\Rv,t)$ are defined in Sec.~\ref{QQEE}, and
\begin{align}
&C^{(n)}(t)=
\langle \Phi(\rvm,t)|\hat{H}^{(e)}(t) | \ \Phi(\rvm,t)\rangle_\rvm   
-i \hbar \langle \Phi(\rvm,t)| \pa_t \Phi(\rvm,t)\rangle_\rvm,\\
&C^{(e)}(t)=
\langle \chi(\Rv,t)|\hat{H}^{(n)}(t) | \ \chi(\Rv,t)\rangle_\Rv   
-i \hbar \langle \chi(\Rv,t)| \pa_t \chi(\Rv,t)\rangle_\Rv.
\end{align}

We, finally, note that the TD constants $C^{(n)}(t)$ and $C^{(e)}(t)$ can be omitted, since they only add phase factors, which do not affect the particle and current densities of the projectile and electrons.

\section{Derivation of Eq.~(\ref{res})}
\label{mainder}

In this Appendix we derive Eq.~(\ref{res}) by expanding in powers of $\Ev$ and $\omega$.

\subsection{1st order in $\Ev$}
\label{1st}
Projectile feels the potential
\begin{equation}
V^{(n)}_{ext}+V^{(n)}_1,
\end{equation}
where $V^{(n)}_1$ is the potential on the projectile from the electronic subsystem. Adopting the operator notations, we can write
for the 1st order change in the projectile's density 
\begin{equation}
n^{(n)}_1=\hat{\chi}^{(n)}_1 (V^{(n)}_{ext}+V^{(n)}_1),
\end{equation}
where $\hat{\chi}^{(n)}_1$ is the linear density response function of the projectile.
Therefore, the 1st order change of the potential acting from the projectile on electrons is
\begin{equation}
V^{(e)}_1=\hat{C} n^{(n)}_1,
\end{equation}
were
\begin{equation}
\hat{C}=-\frac{Z}{|\rv-\Rv|}.
\end{equation}
The 1st order change in the electronic density is
\begin{equation}
n^{(e)}_1=\hat{\chi}^{(e)}_1 V^{(e)}_1,
\end{equation}
where $\hat{\chi}^{(e)}_1$ is the linear density response function of the electronic subsystem. Then, it its turn,
\begin{equation}
V^{(n)}_1=\hat{C} n^{(e)}_1.
\label{987}
\end{equation}

Combining the above equations, we have
\begin{equation}
n^{(n)}_1=\hat{\chi}^{(n)}_1( V^{(n)}_{ext}+\hat{\Pi} n^{(n)}_1),
\label{n1eq}
\end{equation}
where 
\begin{equation}
\hat{\Pi}=\hat{C} \hat{\chi}^{(e)}_1 \hat{C}.
\end{equation}

From Eq.~(\ref{n1eq}) it follows that for $n^{(n)}_1(\Rv,t)$ expansion in $\omega$ starts from $\omega^{-1}$
\begin{equation}
n^{(n)}_1(\Rv,t)=\omega^{-1} n^{(n,-1)}_1(\Rv) \sin(\omega t) + \omega^0 n^{(n,0)}_1(\Rv) \cos(\omega t)  + \dots
\label{n1exp}
\end{equation}
Since $t$ can be large, we only expand in powers of $\omega$ the coefficients at the trigonometric functions. 
The substitution of (\ref{n1exp}) in (\ref{n1eq}) gives
(the action of response functions on real-valued superpositions of cosine and sine waves are listed in Appendix \ref{DRF}) 
\begin{align}
&n^{(n,-1)}_1 \sin(\omega t)=\hat{\chi}^{(n,0)}_1\hat{\Pi}^{(0)} n^{(n,-1)}_1 \sin(\omega t), \label{n-1} \\
&n^{(n,0)}_1 \cos(\omega t) =\hat{\chi}^{(n,0)}_1  V^{(n)}_{ext} \cos(\omega t) +\hat{\chi}^{(n,0)}_1 \hat{\Pi}^{(1)} n^{(n,-1)}_1 \sin(\omega t)+\hat{\chi}^{(n,0)}_1 \hat{\Pi}^{(0)} n^{(n,0)}_1 \cos(\omega t), \label{n-0} 
\end{align}
where we have used the fact that $\hat{\chi}^{(n,1)}_1=0$ due to the discrete energy levels of the projectile at rest.
Equation (\ref{n-1}) has a solution
\begin{equation}
n^{(n,-1)}_1(\Rv)= \av \cdot \nabla n^{(n)}_0(\Rv),
\label{000}
\end{equation}
where $n^{(n)}_0(\Rv)$ is the ground state projectile's density and $\av$ is an arbitrary (as yet) constant vector. Then, from Eq.~(\ref{n-0}),
$\av$ and $n^{(n,0)}_1$ are determined
\begin{comment}
\begin{equation}
[\hat{\chi}^{(n,0)-1}-\hat{\Pi}^{(0)}] n^{(n,0)}_1 \cos(\omega t) =  V^{(n)}_{ext} \cos(\omega t) + \hat{\Pi}^{(1)} n^{(n,-1)}_1 \sin(\omega t),
\end{equation}
\begin{equation}
[\hat{\chi}^{(n,0)-1}-\hat{\Pi}^{(0)}] n^{(n,0)}_1  =  V^{(n)}_{ext} - \left .\pa_\omega \Im \hat{\Pi}^{(1)}(\omega)\right|_{\omega=0} n^{(n,-1)}_1 ,
\end{equation}
\begin{equation}
\langle \nabla_i n^{(n)}_0|V^{(n)}_{ext} -  a_j \left .\pa_\omega \Im \hat{\Pi}(\omega)\right|_{\omega=0} \nabla_j n^{(n)}_0 \rangle=0,
\end{equation} 
\end{comment}
\begin{equation}
\sum\limits_{j=1}^3  a_j \int \nabla_i n^{(n)}_0(\Rv) \left .\pa_\omega \Im \hat{\Pi}(\Rv,\Rv',\omega)\right|_{\omega=0} \nabla'_j n^{(n)}_0(\Rv') d\Rv d\Rv' =E_i,
\label{aeqeq}
\end{equation} 
\begin{equation}
 n^{(n,0)}_1  =  [\hat{\chi}^{(n,0)-1}-\hat{\Pi}^{(0)}]^{-1} [V^{(n)}_{ext} - \left .\pa_\omega \Im \hat{\Pi}(\omega)\right|_{\omega=0} \av \cdot \nabla n^{(n)}_0].
\end{equation}

{\em Derivation of Eq.~(\ref{veqeq}) for the velocity $\vv(t)$}. We can write
\begin{equation}
\vv(t)= \int \Jv^{(n)}(\Rv,t) d\Rv= - \int \Rv (\nabla \cdot \Jv^{(n)}(\Rv,t)) d\Rv=
\pa_t \int \Rv\, n^{(n)}(\Rv,t) d\Rv.
\end{equation}
Since $n^{(n)}_0(\Rv)$ is spherically symmetric, we can write up to the 1st order in $\Ev$
\begin{equation}
\vv(t)= 
\pa_t \int \Rv\, n^{(n)}_1(\Rv,t) d\Rv.
\end{equation}
Then, due to Eq.~(\ref{n1exp}), to the zero-th order in $\omega$
\begin{equation}
\vv(t)= \cos(\omega t)
\int \Rv\, n^{(n,-1)}_1(\Rv) d\Rv,
\end{equation}
or, by virtue of Eq.~(\ref{000}),
\begin{equation}
\vv(t)= - \av \cos(\omega t) 
\int n^{(n)}_0(\Rv) d\Rv= - \av \cos(\omega t).
\label{aaa}
\end{equation}
Equations (\ref{aaa}) and (\ref{aeqeq}) prove Eqs.~(\ref{veq}) and (\ref{veqeq}).

\subsection{2nd order in ${\Ev}$}
Up to the 2nd order, the potential acting on electrons is
\begin{equation}
V^{(e)}=V^{(e)}_1+\hat{C} n^{(n)}_2,
\end{equation}
where $n^{(n)}_2$ is the 2nd order change in the projectile's density. Therefore, the corresponding change in the electron's density is
\begin{equation}
n^{(e)}_2=\hat{\chi}^{(e)}_2 V_1^{(e)} +\hat{\chi}^{(e)}_1 \hat{C} n^{(n)}_2,
\end{equation}
where $\hat{\chi}^{(e)}_2$ is the electronic 2nd-order density response function. Hence, the 2nd order contribution to the potential acting on the projectile is
\begin{equation}
V^{(n)}_2=\hat{C} \hat{\chi}^{(e)}_2 V_1^{(e)} +\hat{\Pi} n^{(n)}_2.
\end{equation}
Up to the 2nd order, the potential acting on the projectile is
\begin{equation}
V^{(n)}=V^{(n)}_{ext}+V^{(n)}_1+V^{(n)}_2.
\end{equation}
Therefore,
\begin{equation}
n^{(n)}_2(\omega) = \hat{\chi}^{(n)}_2 [ V^{(n)}_{ext}(\omega)+V^{(n)}_1(\omega) ]  +\hat{\chi}^{(n)}_1 \hat{C}  \hat{\chi}^{(e)}_2 V^{(e)}_1(\omega)
+ \hat{\chi}^{(n)}_1 \hat{\Pi} n^{(n)}_2(\omega).
\label{n2w}
\end{equation}

Prior to expanding $n^{(n)}_2$ in powers of $\omega$, we separate the cosine, sine, and constant parts in it
\begin{equation}
n^{(n)}_2(\omega) = \cos(2\omega t) n^{(n,c)}(\omega)+\sin(2\omega t) n^{(n,s)}(\omega)+ n^{(n,const)}(\omega).
\label{n2wcs}
\end{equation}
The substitution of Eq.~(\ref{n2wcs}) into (\ref{n2w}) with the use of Eqs.~(\ref{chi2act}) gives a system of two coupled equations (we are not interested in the $n^{(n,const)}$ term, because it does not contribute to Eq.~(\ref{dEkindt2}) )
\begin{equation}
\begin{split}
&n^{(n,c)}_2(\Rv,\omega)=\\
&\frac{1}{2} 
  \hat{\chi}^{(n)}_2(\omega,\omega) [ [-(\Ev\cdot \Rv')+V^{(n,c)}_1(\Rv',\omega)]  [-(\Ev\cdot \Rv'')+V^{(n,c)}_1(\Rv'',\omega)]  \! - \!  V^{(n,s)}_1(\Rv',\omega)  V^{(n,s)}_1(\Rv'',\omega)] \\
 & +\frac{1}{2} \hat{\chi}^{(n)}_1(2\omega) \hat{C}
  \left[  \Re\hat{\chi}^{(e)}_2(\omega,\omega) [  V^{(e,c)}_1(\rv',\omega)  V^{(e,c)}_1(\rv'',\omega)  \! - \!  V^{(e,s)}_1(\rv',\omega)  V^{(e,s)}_1(\rv'',\omega)] \right. \\ &\left.
 - \Im \hat{\chi}^{(e)}_2(\omega,\omega) [V^{(e,c)}_1(\rv',\omega)  V^{(e,s)}_1(\rv'',\omega) \! + \! V^{(e,s)}_1(\rv',\omega)  V^{(e,c)}_1(\rv'',\omega) ]  \right] \\
 &+\hat{\chi}^{(n)}_1(2\omega) [\Re \hat{\Pi}(2 \omega) n^{(n,c)}_2(\omega)-\Im \hat{\Pi}(2 \omega) n^{(n,s)}_2(\omega)],
\end{split}
\label{ccc}
\end{equation}
\begin{equation}
\begin{split}
&n^{(n,s)}_2(\Rv,\omega)=\\
&\frac{1}{2} 
    \hat{\chi}^{(n)}_2(\omega,\omega) [[-(\Ev\cdot \Rv')+V^{(n,c)}_1(\Rv',\omega)]  V^{(n,s)}_1(\Rv'',\omega) \! + \! V^{(n,s)}_1(\Rv',\omega)  [-(\Ev\cdot \Rv'')+V^{(n,c)}_1(\Rv'',\omega)]  ]   \\
&+\frac{1}{2} \hat{\chi}^{(n)}_1(2\omega) \hat{C}
\! \! \left[  \Im \hat{\chi}^{(e)}_2(\omega,\omega) [  V^{(e,c)}_1(\rv',\omega)  V^{(e,c)}_1(\rv'',\omega)  \! - \!  V^{(e,s)}_1(\rv',\omega)  V^{(e,s)}_1(\rv'',\omega)] \right. \\ &\left.
 +  \Re \hat{\chi}^{(e)}_2(\omega,\omega) [V^{(e,c)}_1(\rv',\omega)  V^{(e,s)}_1(\rv'',\omega) \! + \! V^{(e,s)}_1(\rv',\omega)  V^{(e,c)}_1(\rv'',\omega) ]   \right] \\
 &+\hat{\chi}^{(n)}_1(2\omega) [\Im \hat{\Pi}(2 \omega) n^{(n,c)}_2(\omega)+\Re \hat{\Pi}(2 \omega) n^{(n,s)}_2(\omega)].
\end{split}
\label{sss}
\end{equation}

We expand
\begin{equation}
\begin{split}
&n^{(n,c)}(\omega)=\omega^{-2} n^{(n,c,-2)}+\omega^{-1} n^{(n,c,-1)}+\dots,\\
&n^{(n,s)}(\omega)=\omega^{-2} n^{(n,s,-2)}+\omega^{-1} n^{(n,s,-1)}+\dots.
\end{split}
\label{nncs}
\end{equation}
The fact that these expansions start with $\omega^{-2}$ term is a consequence of Eqs.~(\ref{ccc}) and (\ref{sss}). 
Substituting expansions (\ref{nncs}) into Eq.~(\ref{ccc}), we have an equation for $n^{(n,c,-2)}_2$
\begin{equation}
\begin{split}
&n^{(n,c,-2)}_2=
-\frac{1}{2} 
  \hat{\chi}^{(n)}_2(0,0) [   V^{(n,s,-1)}_1(\Rv')  V^{(n,s,-1)}_1(\Rv'')] \\
 & -\frac{1}{2} \hat{\chi}^{(n)}_1(0) \hat{C}
  \hat{\chi}^{(e)}_2(0,0) [     V^{(e,s,-1)}_1(\rv')  V^{(e,s,-1)}_1(\rv'')]  +\hat{\chi}^{(n)}_1(0)  \hat{\Pi}( 0) n^{(n,c,-2)}_2.
\label{eqncm2}
\end{split}
\end{equation}
The solution of this equation is
\begin{equation}
n^{(n,c,-2)}_2(\Rv)=-\frac{1}{4} (\av\cdot \nabla)^2 n^{(n)}_0(\Rv)
\end{equation}
which is a consequence of the fact that Eq.~(\ref{eqncm2}) coincides with the static sum rule describing  the static shift of the whole system by the vector 
$\av$, and $-n^{(n,c,-2)}_2(\Rv)$ appears as the 2nd-order change in the density.
Furthermore,
\begin{align}
&n^{(n,s,-2)}_2=0, \\
&n^{(n,c,-1)}_2=0, \\
\begin{split}
&n^{(n,s,-1)}_2=
\int \hat{\chi}^{(n)}_2(\Rv,\Rv',\Rv'',0,0) [-(\Ev\cdot \Rv')+V^{(n,c,0)}_1(\Rv')]  V^{(n,s,-1)}_1(\Rv'')  d\Rv'' d\Rv'   \\
&-\frac{1}{2} \hat{\chi}^{(n)}_1(0) \hat{C}
\left. \pa_\omega  \Im \hat{\chi}^{(e)}_2(\rv,\rv',\rv'',\omega,\omega) \right|_{\omega=0}   V^{(e,s,-1)}_1(\rv')  V^{(e,s,-1)}_1(\rv'') d\rv'' d\rv' \\
& +   \hat{\chi}^{(n)}_1(0) \hat{C} \int \hat{\chi}^{(e)}_2(\rv,\rv',\rv'',0,0) V^{(e,c,0)}_1(\rv')  V^{(e,s,-1)}_1(\rv'')  d\rv' d\rv'' \\
 &
 +\hat{\chi}^{(n)}_1(0) \left. \pa_\omega \Im \hat{\Pi}(2 \omega)\right|_{\omega=0} n^{(n,c,-2)}_2
 +\hat{\chi}^{(n)}_1(0) \hat{\Pi}(0) n^{(n,s,-1)}_2.
\end{split}
\label{1234}
\end{align}
Equation (\ref{1234}) can be rewritten as
\begin{equation}
\begin{split}
&n^{(n,s,-1)}_2=
a_i \int \hat{\chi}^{(n)}_2(\Rv,\Rv',\Rv'',0,0) [-(\Ev\cdot \Rv')+V^{(n,c,0)}_1(\Rv')] \nabla''_i V^{(n)}_0(\Rv'')  d\Rv'' d\Rv'   \\
&-\frac{1}{2} a_i a_j \hat{\chi}^{(n)}_1(0) \hat{C}
\left. \pa_\omega  \Im \hat{\chi}^{(e)}_2(\rv,\rv',\rv'',\omega,\omega)\right|_{\omega=0}  [\nabla'_i V^{(e)}_0(\rv')] [\nabla''_j V^{(e)}_0(\rv'') d\rv'' d\rv' \\
& +   a_i \hat{\chi}^{(n)}_1(0) \hat{C} \int \hat{\chi}^{(e)}_2(\rv,\rv',\rv'',0,0) V^{(e,c,0)}_1(\rv')  \nabla''_i V^{(e)}_0(\rv'')  d\rv' d\rv'' \\
 &+\hat{\chi}^{(n)}_1(0) \hat{\Pi}(0) n^{(n,s,-1)}_2 +\hat{\chi}^{(n)}_1(0) \left. \pa_\omega \Im \hat{\Pi}(2 \omega)\right|_{\omega=0} n^{(n,c,-2)}_2,
\end{split}
\end{equation}
where we have used Eqs.~(\ref{987}) and (\ref{000}). Using sum rules of Appendix \ref{a2dn}, Eq.~(\ref{sr1}), (\ref{sr2}), and (\ref{sr2}),
we rewrite the last equation as
\begin{equation}
\begin{split}
&n^{(n,s,-1)}_2=\\
&\frac{1}{2} a_i \int [\nabla_i \chi_1^{(n)}(\Rv,\Rv',0)+\nabla'_i \chi_1^{(n)}(\Rv,\Rv',0)] [-(\Ev\cdot \Rv')+V^{(n),c,0}_1(\Rv')]  d\Rv'   \\
&-\frac{1}{4} a_i a_j \hat{\chi}^{(n)}_1(0) \hat{C}
\int \left.\frac{\pa}{\pa \omega} \Im [\nabla_i \chi_1^{(e)}(\rv,\rv',\omega)]\right|_{\omega=0}   [\nabla'_j V^{(e)}_0(\rv')]  d\rv'  \\
&-\frac{1}{4} a_i a_j \hat{\chi}^{(n)}_1(0) \hat{C}
\int \left.\frac{\pa}{\pa \omega} \Im [\nabla_j \chi_1^{(e)}(\rv,\rv',\omega)]\right|_{\omega=0}   [\nabla'_i V^{(e)}_0(\rv')]  d\rv' \\
&-\frac{1}{4} a_i a_j \hat{\chi}^{(n)}_1(0) \hat{C}
\int \left.\frac{\pa}{\pa \omega} \Im \nabla'_i \chi_1^{(e)}(\rv,\rv',2 \omega)]\right|_{\omega=0}   [\nabla'_j V^{(e)}_0(\rv')]  d\rv'\\
& +  \frac{1}{2} a_i \hat{\chi}^{(n)}_1(0) \hat{C} \int [\nabla_i \chi_1^{(e)}(\rv,\rv',0)+\nabla'_i \chi_1^{(e)}(\rv,\rv',0)] V^{(e,c,0)}_1(\rv')    d\rv'  \\
 &+\hat{\chi}^{(n)}_1(0) \hat{\Pi}(0) n^{(n),s,-1}_2 +\hat{\chi}^{(n)}_1(0) \left. \pa_\omega \Im \hat{\Pi}(2 \omega)\right|_{\omega=0} n^{(n,c,-2)}_2.
\end{split}
\label{5643}
\end{equation}
We note, that no second order response functions are present in Eq.~(\ref{5643}) any more.
The sequence of the following transformations, leads us to an explicit result (\ref{good}) for $n^{(n,s,-1)}_2$
\begin{equation}
\begin{split}
&[1-\hat{\chi}^{(n)}_1(0) \hat{\Pi}(0)]n^{(n,s,-1)}_2=\\
&\frac{1}{2} a_i \int [\nabla_i \chi_1^{(n)}(\Rv,\Rv',0)+\nabla'_i \chi_1^{(n)}(\Rv,\Rv',0)] [-(\Ev\cdot \Rv')+V^{(n,c,0)}_1(\Rv')]  d\Rv'   \\
&-\frac{1}{2} a_i a_j \hat{\chi}^{(n)}_1(0) \hat{C}
\int \left.\frac{\pa}{\pa \omega} \Im [\nabla_i \chi_1^{(e)}(\rv,\rv',\omega)]\right|_{\omega=0}   [\nabla'_j V^{(e)}_0(\rv')]  d\rv' \\
&-\frac{1}{4} a_i a_j \hat{\chi}^{(n)}_1(0) \hat{C}
\int \left.\frac{\pa}{\pa \omega} \Im \nabla'_i \chi_1^{(e)}(\rv,\rv',2 \omega)]\right|_{\omega=0}   [\nabla'_j V^{(e)}_0(\rv')]  d\rv'\\
& +  \frac{1}{2} a_i \hat{\chi}^{(n)}_1(0) \hat{C} \int [\nabla_i \chi_1^{(e)}(\rv,\rv',0)+\nabla'_i \chi_1^{(e)}(\rv,\rv',0)] V^{(e,c,0)}_1(\rv')    d\rv'  
+\hat{\chi}^{(n)}_1(0) \left. \pa_\omega \Im \hat{\Pi}(2 \omega)\right|_{\omega=0} n^{(n,c,-2)}_2
\end{split}
\end{equation}
\begin{equation}
-(\Ev\cdot \Rv)+V^{(n,c,0)}_1(\Rv)=[\chi^{(n)}_1(0)]^{-1} n^{(n,c,0)}
\end{equation}
\begin{equation}
\begin{split}
&[1-\hat{\chi}^{(n)}_1(0) \hat{\Pi}(0)]n^{(n,s,-1)}_2=\\
&\frac{1}{2} a_i \int \nabla_i \chi_1^{(n)}(\Rv,\Rv',0) [-(\Ev\cdot \Rv')+V^{(n,c,0)}_1(\Rv')]  d\Rv'\\
&+\frac{1}{2} a_i \int \nabla'_i \chi_1^{(n)}(\Rv,\Rv',0) [V^{(n,c,0)}_1(\Rv')]  d\Rv'   \\
&-\frac{1}{2} a_i a_j \hat{\chi}^{(n)}_1(0) \hat{C}
\int \left.\frac{\pa}{\pa \omega} \Im [\nabla_i \chi_1^{(e)}(\rv,\rv',\omega)]\right|_{\omega=0}   [\nabla'_j V^{(e)}_0(\rv')]  d\rv' \\
&-\frac{1}{4} a_i a_j \hat{\chi}^{(n)}_1(0) \hat{C}
\int \left.\frac{\pa}{\pa \omega} \Im \nabla'_i \chi_1^{(e)}(\rv,\rv',2 \omega)]\right|_{\omega=0}   [\nabla'_j V^{(e)}_0(\rv')]  d\rv'\\
& +  \frac{1}{2} a_i \hat{\chi}^{(n)}_1(0) \hat{C} \int [\nabla_i \chi_1^{(e)}(\rv,\rv',0)+\nabla'_i \chi_1^{(e)}(\rv,\rv',0)] V^{(e,c,0)}_1(\rv')    d\rv'
+\hat{\chi}^{(n)}_1(0) \left. \pa_\omega \Im \hat{\Pi}(2 \omega)\right|_{\omega=0} n^{(n,c,-2)}_2  
\end{split}
\end{equation}
\begin{equation}
\begin{split}
&[1-\hat{\chi}^{(n)}_1(0) \hat{\Pi}(0)]n^{(n,s,-1)}_2=
[1-\hat{\chi}^{(n)}_1(0) \hat{\Pi}(0)] \frac{1}{2} a_i \nabla_i n^{(n,c,0)}_1 \\
&+\frac{1}{2} a_i \int \nabla'_i \chi_1^{(n)}(\Rv,\Rv',0) [V^{(n),c,0}_1(\Rv')]  d\Rv'   \\
&-\frac{1}{2} a_i a_j \hat{\chi}^{(n)}_1(0) \hat{C}
\int \left.\frac{\pa}{\pa \omega} \Im [\nabla_i \chi_1^{(e)}(\rv,\rv',\omega)]\right|_{\omega=0}   [\nabla'_j V^{(e)}_0(\rv')]  d\rv' \\
&-\frac{1}{4} a_i a_j \hat{\chi}^{(n)}_1(0) \hat{C}
\int \left.\frac{\pa}{\pa \omega} \Im \nabla'_i \chi_1^{(e)}(\rv,\rv',2 \omega)]\right|_{\omega=0}   [\nabla'_j V^{(e)}_0(\rv')]  d\rv'\\
& +  \frac{1}{2} a_i \hat{\chi}^{(n)}_1(0) \hat{C} \int [\nabla_i \chi_1^{(e)}(\rv,\rv',0)] V^{(e,c,0)}_1(\rv')    d\rv' 
+\hat{\chi}^{(n)}_1(0) \left. \pa_\omega \Im \hat{\Pi}(2 \omega)\right|_{\omega=0} n^{(n,c,-2)}_2
\end{split}
\end{equation}
\begin{equation}
\begin{split}
&[1-\hat{\chi}^{(n)}_1(0) \hat{\Pi}(0)]n^{(n,s,-1)}_2=
[1-\hat{\chi}^{(n)}_1(0) \hat{\Pi}(0)] \frac{1}{2} a_i \nabla_i n^{(n,c,0)}_1 \\
&-\frac{1}{2} a_i  \chi_1^{(n)}(0) \nabla_i [\hat{\Pi}(0) n^{(n,c,0)}_1- \left. \pa_\omega \Im  \hat{\Pi}(\omega)\right|_{\omega=0}  n^{(n,s,-1)}_1]     \\
&-\frac{1}{2} a_i a_j \hat{\chi}^{(n)}_1(0) 
\int \left.\frac{\pa}{\pa \omega} \Im [\nabla_i \Pi(\rv,\rv',\omega)]\right|_{\omega=0}   [\nabla'_j n^{(n)}_0(\rv')]  d\rv' \\
&+\frac{1}{2} a_i a_j \hat{\chi}^{(n)}_1(0) 
\int \left.\frac{\pa}{\pa \omega} \Im \Pi(\rv,\rv', \omega)]\right|_{\omega=0}   [\nabla'_i  \nabla'_j n^{(0)}_0(\rv')]  d\rv'\\
& +  \frac{1}{2} a_i \hat{\chi}^{(n)}_1(0)  \nabla_i \Pi(\rv,\rv',0) n^{(n,c,0)}_1 
+\hat{\chi}^{(n)}_1(0) \left. \pa_\omega \Im \hat{\Pi}(2 \omega)\right|_{\omega=0} n^{(n,c,-2)}_2
\end{split}
\end{equation}
\begin{equation}
\begin{split}
&[1-\hat{\chi}^{(n)}_1(0) \hat{\Pi}(0)]n^{(n,s,-1)}_2=
[1-\hat{\chi}^{(n)}_1(0) \hat{\Pi}(0)] \frac{1}{2} a_i \nabla_i n^{(n,c,0)}_1 \\
&+\frac{1}{2} a_i  a_j \chi_1^{(n)}(0) \nabla_i  \left. \pa_\omega \Im  \hat{\Pi}(\omega)\right|_{\omega=0} \nabla_j  n^{(n)}_0     \\
&-\frac{1}{2} a_i a_j \hat{\chi}^{(n)}_1(0) 
\int \left.\frac{\pa}{\pa \omega} \Im [\nabla_i \Pi(\rv,\rv',\omega)]\right|_{\omega=0}   [\nabla'_j n^{(n)}_0(\rv')]  d\rv' \\
&+\frac{1}{2} a_i a_j \hat{\chi}^{(n)}_1(0) 
\int \left.\frac{\pa}{\pa \omega} \Im \Pi(\rv,\rv', \omega)]\right|_{\omega=0}   [\nabla'_i  \nabla'_j n^{(0)}_0(\rv')]  d\rv'
+\hat{\chi}^{(n)}_1(0) \left. \pa_\omega \Im \hat{\Pi}(2 \omega)\right|_{\omega=0} n^{(n,c,-2)}_2
\end{split}
\end{equation}
\begin{equation}
\begin{split}
&[1-\hat{\chi}^{(n)}_1(0) \hat{\Pi}(0)]n^{(n,s,-1)}_2=
[1-\hat{\chi}^{(n)}_1(0) \hat{\Pi}(0)] \frac{1}{2} a_i \nabla_i n^{(n,c,0)}_1     \\
&+\frac{1}{2} a_i a_j \hat{\chi}^{(n)}_1(0) 
\int \left.\frac{\pa}{\pa \omega} \Im \Pi(\Rv,\Rv', \omega)]\right|_{\omega=0}   [\nabla'_i  \nabla'_j n^{(0)}_0(\Rv')  d\rv'
+\hat{\chi}^{(n)}_1(0) \left. \pa_\omega \Im \hat{\Pi}(2 \omega)\right|_{\omega=0} n^{(n,c,-2)}_2
\end{split}
\end{equation}
\begin{equation}
\begin{split}
&[[\hat{\chi}^{(n)}_1(0)]^{-1}- \hat{\Pi}(0)]n^{(n,s,-1)}_2=
[[\hat{\chi}^{(n)}_1(0)]^{-1}- \hat{\Pi}(0)] \frac{1}{2} a_i \nabla_i n^{(n,c,0)}_1     \\
&+\frac{1}{2} a_i a_j 
 \left.\frac{\pa}{\pa \omega} \Im \Pi( \omega)]\right|_{\omega=0}   \nabla_i  \nabla_j n^{(n)}_0
 +\hat{\chi}^{(n)}_1(0) \left. \pa_\omega \Im \hat{\Pi}(2 \omega)\right|_{\omega=0} n^{(n,c,-2)}_2.
\end{split}
\end{equation}
\begin{equation}
\begin{split}
&[[\hat{\chi}^{(n)}_1(0)]^{-1}- \hat{\Pi}(0)]n^{(n,s,-1)}_2=[[\hat{\chi}^{(n)}_1(0)]^{-1}- \hat{\Pi}(0)] \frac{1}{2} a_i \nabla_i n^{(n,c,0)}
\end{split}
\end{equation}

\begin{equation}
n^{(n,s,-1)}_2=\frac{1}{2} a_i \nabla_i n^{(n,c,0)}_1.
\label{good}
\end{equation}

We rewrite Eq.~(\ref{dEkindt2}) as
\begin{equation}
\begin{split}
&\left(\frac{d E_{kin}}{d t}\right)_2=
2 \omega \sin(2\omega t) \int V^{(n)}_0(\Rv)  n^{(n,c)}_2(\Rv) d\Rv 
-2 \omega \cos(2\omega t) \int V^{(n)}_0(\Rv)  n^{(n,s)}_2(\Rv) d\Rv \\
&+\omega \int  [V^{(n,c)}_1(\Rv)\cos(\omega t)+V^{(n,s)}_1(\Rv)\sin(\omega t)  [n^{(n,c)}_1(\Rv) \sin(\omega t)-n^{(n,s)}_1(\Rv) \cos(\omega t)]  d\Rv.
\end{split}
\end{equation}
\begin{equation}
\begin{split}
&\left(\frac{d E_{kin}}{d t}\right)_2=
2 \omega \sin(2\omega t) \int V^{(n)}_0(\Rv)  n^{(n,c)}_2(\Rv) d\Rv 
-2 \omega \cos(2\omega t) \int V^{(n)}_0(\Rv)  n^{(n,s)}_2(\Rv) d\Rv \\
&+\frac{1}{2}\omega \sin(2\omega t) \int   [V^{(n,c)}_1(\Rv)n^{(n,c)}_1(\Rv)-V^{(n,s)}_1(\Rv) n^{(n,s)}_1(\Rv)] d\Rv  \\  
&+\frac{1}{2}\omega \int [V^{(n,s)}_1(\Rv) n^{(n,c)}_1(\Rv)-V^{(n,c)}_1(\Rv) n^{(n,s)}_1(\Rv)] d\Rv \\
&- \frac{1}{2}\omega \cos(2\omega t) \int [V^{(n,s)}_1(\Rv) n^{(n,c)}_1(\Rv) 
+V^{(n,c)}_1(\Rv) n^{(n,s)}_1(\Rv) ] d\Rv
\end{split}
\end{equation}
\begin{equation}
\begin{split}
&\left(\frac{d E_{kin}}{d t}\right)^{(-1)}_2=
 -\frac{1}{2}  \sin(2\omega t) \int V^{(n)}_0(\Rv)  (\av\cdot \nabla)^2 n^{(n)}_0(\Rv) d\Rv  \\
&-\frac{1}{2} \sin(2\omega t) \int   (\av\cdot \nabla) V^{(n)}_0(\Rv)(\av\cdot \nabla) n_0(\Rv) d\Rv  =0,
\end{split}
\end{equation}
where, as always, the lower and the upper index denotes the order in $\Ev$ and in $\omega$, respectively.
\begin{equation}
\begin{split}
&\left(\frac{d E_{kin}}{d t}\right)^{(0)}_2=
-2  \cos(2\omega t) \int V^{(n)}_0(\Rv)  n^{(n,s,-1)}_2(\Rv) d\Rv\\  
&-\frac{1}{2} \int V^{(n,c,0)}_1(\Rv)(\av\cdot \nabla) n^{(n)}_0(\Rv) d\Rv 
+\frac{1}{2} \int [(\av\cdot \nabla) V^{(n)}_0(\Rv)] n^{(n,c,0)}_1(\Rv) d\Rv\\
&- \frac{1}{2} \cos(2\omega t) \int [(\av\cdot \nabla) V^{(n)}_0(\Rv)] n^{(n,c,0)}_1(\Rv) 
 d\Rv
- \frac{1}{2} \cos(2\omega t) \int V^{(n,c,0)}_1(\Rv) (\av\cdot \nabla) n^{(n)}_0(\Rv)  d\Rv
\end{split}
\end{equation}
\begin{equation}
\begin{split}
&\left(\frac{d E_{kin}}{d t}\right)^{(0)}_2=
-2  \cos(2\omega t) \int V^{(n)}_0(\Rv)  n^{(n,s,-1)}_2(\Rv) d\Rv \\
&+\sin^2(\omega t) \int [(\av\cdot \nabla) V^{(n)}_0(\Rv)] n^{(n,c,0)}_1(\Rv) 
 d\Rv
-  \cos^2(\omega t) \int V^{(n,c,0)}_1(\Rv) (\av\cdot \nabla) n^{(n)}_0(\Rv)  d\Rv
\end{split}
\end{equation}
\begin{equation}
\begin{split}
&\left(\frac{d E_{kin}}{d t}\right)^{(0)}_2=
-2  \cos(2\omega t) \int V^{(n)}_0(\Rv)  n^{(n,s,-1)}_2(\Rv) d\Rv \\
&+\sin^2(\omega t) \int [(\av\cdot \nabla) V^{(n)}_0(\Rv)] n^{(n,c,0)}_1(\Rv) 
 d\Rv\\
&-  \cos^2(\omega t) \int [(\Ev\cdot\Rv) + \left .\pa_\omega \Im \hat{\Pi}(\omega)\right|_{\omega=0} \av \cdot \nabla n_0
+[\chi^{(n)}_1(0)]^{-1} n^{(n,c,0)}_1] (\av\cdot \nabla) n^{(n)}_0(\Rv)]  d\Rv\\
&+  \cos^2(\omega t) \int [(\av\cdot \nabla) n^{(n)}_0(\Rv)] [ \left. \pa_\omega \Im  \hat{\Pi}(,\Rv,\Rv',\omega)\right|_{\omega=0}  [(\av\cdot \nabla) n^{(n)}_0(\Rv')]   d\Rv d\Rv'
\end{split}
\end{equation}
\begin{equation}
\begin{split}
&\left(\frac{d E_{kin}}{d t}\right)^{(0)}_2=
-2  \cos(2\omega t) \int V^{(n)}_0(\Rv)  n^{(n,s,-1)}_2(\Rv) d\Rv \\
&-\cos(2\omega t) \int [(\av\cdot \nabla) V^{(n)}_0(\Rv)] n^{(n,c,0)}_1(\Rv) 
 d\Rv
+ \cos^2(\omega t) (\Ev\cdot\av) 
\end{split}
\end{equation}
\begin{equation}
\begin{split}
&\left(\frac{d E_{kin}}{d t}\right)^{(0)}_2=
-2  \cos(2\omega t) \int V^{(n)}_0(\Rv)  n^{(n,s,-1)}_2(\Rv) d\Rv \\
&-\cos(2\omega t) \int [(\av\cdot \nabla) V^{(n)}_0(\Rv)] n^{(n,c,0)}_1(\Rv) 
 d\Rv\\
&+ \cos^2(\omega t)  a_i a_j \int \nabla_i n_0(\Rv) \left .\pa_\omega \Im \hat{\Pi}(\Rv,\Rv',\omega)\right|_{\omega=0} \nabla'_j n_0(\Rv') d\Rv d\Rv'
\end{split}
\end{equation}
\begin{equation}
\begin{split}
&\left(\frac{d E_{kin}}{d t}\right)^{(0)}_2=
-2  \cos(2\omega t) \int V^{(n)}_0(\Rv)  n^{(n,s,-1)}_2(\Rv) d\Rv \\
&-\cos(2\omega t) \int [(\av\cdot \nabla) V^{(n)}_0(\Rv)] n^{(n,c,0)}_1(\Rv) 
 d\Rv\\
&+ \cos^2(\omega t)  a_i a_j \int \nabla_i n_0(\Rv) \left .\pa_\omega \Im \hat{\Pi}(\Rv,\Rv',\omega)\right|_{\omega=0} \nabla'_j n_0(\Rv') d\Rv d\Rv'
\end{split}
\end{equation}
\begin{equation}
\begin{split}
&\left(\frac{d E_{kin}}{d t}\right)^{(0)}_2=
-  \cos(2\omega t) a_i\int V^{(n)}_0(\Rv)   \nabla_i n^{(n,c,0)}(\Rv) d\Rv \\
&-\cos(2\omega t) \int [(\av\cdot \nabla) V^{(n)}_0(\Rv)] n^{(n,c,0)}_1(\Rv) 
 d\Rv\\
&+ \cos^2(\omega t)  a_i a_j \int \nabla_i n_0(\Rv) \left .\pa_\omega \Im \hat{\Pi}(\Rv,\Rv',\omega)\right|_{\omega=0} \nabla'_j n_0(\Rv') d\Rv d\Rv'
\end{split}
\end{equation}
\begin{equation}
\begin{split}
&\left(\frac{d E_{kin}}{d t}\right)^{(0)}_2=
 \cos^2(\omega t)   \int (\av\cdot \nabla_i) n_0(\Rv) \left .\pa_\omega \Im \hat{\Pi}(\Rv,\Rv',\omega)\right|_{\omega=0} (\av\cdot \nabla'_j) n_0(\Rv') d\Rv d\Rv'.
\end{split}
\label{verygood}
\end{equation}
Equation (\ref{verygood}) together with Eqs.~(\ref{veq})-(\ref{Qprel})  proves Eq.~(\ref{res}).

\section{Density response functions}
\label{DRF}

\subsection{1st order}

The first order response function is defined as
\begin{equation}
\delta n_1(\rv,t)= \int \chi_1(\rv,\rv',t-t') \delta v(\rv,t') d\rv' d t'.
\end{equation}
It acts on a superposition of cosine and sine waves as
\begin{equation}
\begin{split}
&\int \chi_1(\rv,\rv',t-t') [ P(\rv') \cos(\omega t') +Q(\rv') \sin(\omega t')] d t' d\rv' \\ 
&=
\cos(\omega t) \int [\Re [\chi_1(\rv,\rv',\omega)] P(\rv') -\Im [\chi_1(\rv,\rv',\omega)] Q(\rv')] d\rv'\\ 
&+
\sin(\omega t) \int [\Im [\chi_1(\rv,\rv',\omega)] P(\rv') +\Re [\chi_1(\rv,\rv',\omega)] Q(\rv')] d\rv',
\end{split}
\end{equation}
where $\chi_1(\rv,\rv',\omega)$ is Fourier transform of $\chi_1(\rv,\rv',t-t')$.

{\em The Lehmann representation} holds
\begin{equation}
\chi_1(\rv,\rv',\omega)=\langle 0|\hat{n}(\rv)|n\rangle \langle n|\hat{n}(\rv')|0\rangle
\left[ \frac{1}{E_0-E_n-\omega-i\eta} -\frac{1}{E_n-E_0-\omega-i\eta}\right],
\end{equation}
where $|n\rangle$ and $E_n$ are many-body wave-functions and energies, respectively. Summation is implied over repeated state's symbols.

\subsection{2nd order}
\label{a2dn}

The second order response function is defined as
\begin{equation}
\delta n_2(\rv,t)= \int \chi_2(\rv,\rv',\rv'',t-t',t-t'') \delta v(\rv,t') \delta v(\rv'',t'') d\rv' d\rv'' d t' d t''.
\end{equation}
It acts on a superposition of cosine and sine waves as
\begin{equation}
\begin{split}
&\int d t' d t'' \chi_2(\rv,\rv',\rv'',t-t',t-t'') [ P(\rv') \cos(\omega t') +Q(\rv') \sin(\omega t')]  \\
&\times [ P(\rv'') \cos(\omega t'') +Q(\rv'') \sin(\omega t'')] 
 = \frac{\cos(2\omega t)}{2} C(\rv,t)+\frac{\sin(2\omega t)}{2} S(\rv,t)+\frac{1}{2} D(\rv,t),
\end{split}
\label{chi2act}
\end{equation}
\begin{align*}
&C(\rv,t)= \\
&\int \! \! \left[  \Re \chi_2(\rv,\rv',\rv'', \omega,\omega) [P(\rv') P(\rv'') \! - \! Q(\rv') Q(\rv'')]
\! - \! \Im \chi_2(\rv,\rv',\rv'', \omega,\omega) [P(\rv') Q(\rv'') \! + \! Q(\rv') P(\rv'')]  \right] \! d\rv' d\rv'',\\
&S(\rv,t)= \\
&\int \! \! \left[  \Im \chi_2(\rv,\rv',\rv'', \omega,\omega) [P(\rv') P(\rv'') \! - \! Q(\rv') Q(\rv'')]
\! + \! \Re \chi_2(\rv,\rv',\rv'', \omega,\omega) [P(\rv') Q(\rv'') \! + \! Q(\rv') P(\rv'')]  \right] \! d\rv' d\rv'', \\
&D(\rv,t) = \\
&\int \! \! \left[  \Re \chi_2(\rv,\rv',\rv'', \omega,-\omega) [P(\rv') P(\rv'') \! + \! Q(\rv') Q(\rv'')]
\! + \! \Im \chi_2(\rv,\rv',\rv'', \omega,-\omega) [Q(\rv') P(\rv'') \! - \! P(\rv') Q(\rv'')]   \right] \! d\rv' d\rv'',
\end{align*}
where $\chi_2(\rv,\rv',\rv'',\omega,\omega_1)$ is a double Fourier transform of $\chi_2(\rv,\rv',\rv'',t-t',t-t'')$.

{\em The Lehmann representation} holds
\begin{equation}
\begin{split}
&\chi_2(\rv,\rv',\rv'',\omega_1,\omega_2) = \frac{1}{2} \times \\
&  \left[ \frac{  
\langle 0|\hat{n}(\rv)| m\rangle \langle m|\hat{n}(\rv')| n\rangle 
\langle n|\hat{n}(\rv'')| 0\rangle}{ (E_n-E_0-\omega_2-i\eta)  (E_m-E_0-\omega_1-\omega_2-i\eta)} \right.
\\
&-  \frac{
\langle n|\hat{n}(\rv)| m\rangle \langle m|\hat{n}(\rv')| 0\rangle  
\langle 0|\hat{n}(\rv'')| n\rangle}{ (E_0-E_n-\omega_2-i\eta) (E_m-E_n-\omega_1-\omega_2-i\eta)}
 \\
&-  \frac{ 
\langle n|\hat{n}(\rv)| m\rangle \langle m|\hat{n}(\rv'')| 0\rangle
 \langle 0|\hat{n}(\rv')| n\rangle}{ (E_m-E_0-\omega_2-i\eta)  (E_m-E_n-\omega_1-\omega_2-i\eta) }\\
&\left. +  \frac{
\langle n|\hat{n}(\rv)| 0\rangle \langle 0|\hat{n}(\rv'')| m\rangle  
 \langle m|\hat{n}(\rv')| n\rangle}{ (E_0-E_m-\omega_2-i\eta) (E_0-E_n-\omega_1-\omega_2-i\eta)} +( \rv' \leftrightarrow \rv'', \omega_1  \leftrightarrow \omega_2) \right]
\end{split}
\end{equation}

\newpage

{\em Sum rules.}

\begin{equation}
\begin{split}
&2 \int \chi_2(\rv,\rv',\rv'',\omega_1,\omega_2) \nabla_i V_0(\rv'') d\rv''= \\
&  - \frac{ \omega_2 
\langle 0|\hat{n}(\rv)| m\rangle \langle m|\hat{n}(\rv')| n\rangle 
\langle n|\nabla_i| 0\rangle}{ (E_n-E_0-\omega_2-i\eta)  (E_m-E_0-\omega_1-\omega_2-i\eta)}
-   \frac{\omega_2
\langle n|\hat{n}(\rv)| m\rangle \langle m|\hat{n}(\rv')| 0\rangle  
\langle n|\nabla_i| 0\rangle}{ (E_0-E_n-\omega_2-i\eta) (E_m-E_n-\omega_1-\omega_2-i\eta)}
 \\
&+  \frac{ \omega_2
\langle n|\hat{n}(\rv)| m\rangle 
 \langle 0|\hat{n}(\rv')| m\rangle \langle n|\nabla_i| 0\rangle}{ (E_n-E_0-\omega_2-i\eta)  (E_n-E_m-\omega_1-\omega_2-i\eta) }
+   \frac{\omega_2
\langle m|\hat{n}(\rv)| 0\rangle   
 \langle m|\hat{n}(\rv')| n\rangle \langle n|\nabla_i| 0\rangle}{ (E_0-E_n-\omega_2-i\eta) (E_0-E_m-\omega_1-\omega_2-i\eta)}    \\
&+\nabla_i \chi_1(\rv,\rv',\omega_1)+\nabla'_i \chi_1(\rv,\rv',\omega_1+\omega_2)
 \\
& -\frac{  \omega_2
\langle 0|\hat{n}(\rv)| m\rangle  
\langle n|\hat{n}(\rv')| 0\rangle \langle m|\nabla| n\rangle  }{ (E_n-E_0-\omega_1-i\eta)  (E_m-E_0-\omega_1-\omega_2-i\eta)}
+  \frac{\omega_2
\langle n|\hat{n}(\rv)| m\rangle  
\langle m|\hat{n}(\rv')| 0\rangle \langle n|\nabla| 0\rangle  }{ (E_0-E_m-\omega_1-i\eta) (E_n-E_m-\omega_1-\omega_2-i\eta)}
 \\
&-   \frac{ \omega_2
\langle n|\hat{n}(\rv)| m\rangle 
  \langle m|\hat{n}(\rv')| 0\rangle \langle n|\nabla| 0\rangle }{ (E_m-E_0-\omega_1-i\eta)  (E_m-E_n-\omega_1-\omega_2-i\eta) }
+  \frac{\omega_2
\langle m|\hat{n}(\rv)| 0\rangle   
 \langle n|\hat{n}(\rv')| 0\rangle \langle m|\nabla| n\rangle }{ (E_0-E_n-\omega_1-i\eta) (E_0-E_m-\omega_1-\omega_2-i\eta)} 
\end{split}
\end{equation}
In particular,
\begin{equation}
\begin{split}
2 \int \chi_2(\rv,\rv',\rv'',0,0) \nabla_i V_0(\rv'') d\rv''= 
\nabla_i \chi_1(\rv,\rv',0)+\nabla'_i \chi_1(\rv,\rv',0)
\end{split}
\label{sr1}
\end{equation}

\begin{equation}
\begin{split}
&2 \int \chi_2(\rv,\rv',\rv'',\omega_1,\omega_2) \nabla_i V_0(\rv'') d\rv'' \nabla_j' V_0(\rv')  d\rv'= 
-\nabla_i \nabla_j n_0(\rv)   \\
&  +\frac{  \omega_1\omega_2 
\langle 0|\hat{n}(\rv)| m\rangle \langle m|\nabla_j | n\rangle 
\langle n|\nabla_i| 0\rangle}{ (E_n-E_0-\omega_2-i\eta)  (E_m-E_0-\omega_1-\omega_2-i\eta)}
 +  \frac{ \omega_1 \omega_2 
\langle n|\hat{n}(\rv)| m\rangle \langle m|\nabla_j| 0\rangle  
\langle n|\nabla_i| 0\rangle}{ (E_0-E_n-\omega_2-i\eta) (E_m-E_n-\omega_1-\omega_2-i\eta)}
   \\
&+  \frac{\omega_1 \omega_2
\langle n|\hat{n}(\rv)| m\rangle 
 \langle m|\nabla_j | 0\rangle \langle n|\nabla_i| 0\rangle}{ (E_n-E_0-\omega_2-i\eta)  (E_n-E_m-\omega_1-\omega_2-i\eta) }
+  \frac{ \omega_1 \omega_2 
\langle m|\hat{n}(\rv)| 0\rangle   
 \langle m| \nabla_j | n\rangle \langle n|\nabla_i| 0\rangle}{ (E_0-E_n-\omega_2-i\eta) (E_0-E_m-\omega_1-\omega_2-i\eta)}   \\
&+\int [\nabla_i \chi_1(\rv,\rv',\omega_1)] \nabla_j' V_0(\rv') d\rv'
 +\int [\nabla_j \chi_1(\rv,\rv',\omega_2)]\nabla_i' V_0(\rv') d\rv'
+\int [\nabla'_i \chi_1(\rv,\rv',\omega_1+\omega_2)] \nabla_j' V_0(\rv') d\rv'
 \\
& +\frac{ \omega_1 \omega_2
\langle 0|\hat{n}(\rv)| n\rangle  
\langle m| \nabla_j | 0\rangle \langle n|\nabla_i| m\rangle  }{ (E_m-E_0-\omega_1-i\eta)  (E_n-E_0-\omega_1-\omega_2-i\eta)}
+  \frac{\omega_1 \omega_2
\langle n|\hat{n}(\rv)| m\rangle  
\langle m|\nabla_j | 0\rangle \langle n|\nabla| 0\rangle  }{ (E_0-E_m-\omega_1-i\eta) (E_n-E_m-\omega_1-\omega_2-i\eta)}
 \\
&+ \frac{\omega_1  \omega_2
\langle n|\hat{n}(\rv)| m\rangle 
  \langle m|\nabla_j| 0\rangle \langle n|\nabla| 0\rangle }{ (E_m-E_0-\omega_1-i\eta)  (E_m-E_n-\omega_1-\omega_2-i\eta) }
+ \frac{\omega_1  \omega_2
\langle n|\hat{n}(\rv)| 0\rangle   
 \langle m|\nabla_j | 0\rangle \langle n|\nabla_i| m\rangle }{ (E_0-E_m-\omega_1-i\eta) (E_0-E_n-\omega_1-\omega_2-i\eta)} 
\end{split}
\end{equation}
In particular
\begin{equation}
\begin{split}
&2 \int \chi_2(\rv,\rv',\rv'',0,0) \nabla_i V_0(\rv'') d\rv'' \nabla_j' V_0(\rv')  d\rv'= 
 -\nabla_i \nabla_j n_0(\rv)   \\
&+\int [\nabla_i \chi_1(\rv,\rv',0)] \nabla_j' V_0(\rv') d\rv'
 +\int [\nabla_j \chi_1(\rv,\rv',0)]\nabla_i' V_0(\rv') d\rv'
+\int [\nabla'_i \chi_1(\rv,\rv',0)] \nabla_j' V_0(\rv') d\rv',
\label{sr2}
\end{split}
\end{equation}
\begin{equation}
\begin{split}
&2 \int \left. \pa_\omega \chi_2(\rv,\rv',\rv'',\omega,\omega)\right|_{\omega=0} \nabla_i V_0(\rv'') d\rv'' \nabla_j' V_0(\rv')  d\rv'= \\
& \left. \pa_\omega \left\{\int [\nabla_i \chi_1(\rv,\rv',\omega)] \nabla_j' V_0(\rv') d\rv'
 +\int [\nabla_j \chi_1(\rv,\rv',\omega)]\nabla_i' V_0(\rv') d\rv'
+\int [\nabla'_i \chi_1(\rv,\rv',2 \omega)] \nabla_j' V_0(\rv') d\rv' \right\} \right|_{\omega=0}.
\end{split}
\label{sr3}
\end{equation}

%\bibliography{ref}

%aipnum4-2.bst 2019-01-14 (MD) hand-edited version of apsrev4-1.bst

\end{document}